\documentclass[superscriptaddress,twocolumn,aps,prx,10pt]{revtex4-1}

\usepackage{times}
\usepackage{amsmath}
\usepackage{amsfonts}
\usepackage{amssymb}
\usepackage{bm}
\usepackage{graphicx}
\usepackage{color}
\usepackage{float}

\begin{document}
\title{Spontaneous membrane formation and self-encapsulation of active rods in an inhomogeneous motility field}

\author{Jens Grauer}
\affiliation{Institute for Theoretical Physics II: Soft Matter, Heinrich-Heine University D\"{u}sseldorf,
             Universit\"{a}tsstra{\ss}e 1, 40225 D\"{u}sseldorf, Germany}
\author{Hartmut L\"{o}wen}
\affiliation{Institute for Theoretical Physics II: Soft Matter, Heinrich-Heine University D\"{u}sseldorf,
             Universit\"{a}tsstra{\ss}e 1, 40225 D\"{u}sseldorf, Germany}
\author{Liesbeth M.~C.~Janssen}
\email[Electronic mail: ]{L.M.C.Janssen@tue.nl}
\affiliation{Institute for Theoretical Physics II: Soft Matter, Heinrich-Heine University D\"{u}sseldorf,
             Universit\"{a}tsstra{\ss}e 1, 40225 D\"{u}sseldorf, Germany}
\affiliation{Theory of Polymers and Soft Matter, Department of Applied Physics, Eindhoven University of Technology,
             P.O. Box 513, 5600MB Eindhoven, The Netherlands}

\date{\today}

\begin{abstract}
We study the collective dynamics of self-propelled rods in an inhomogeneous
motility field. At the interface between two regions of constant but different
motility, a smectic rod layer is spontaneously created through aligning
interactions between the active rods, reminiscent of an artificial, semi-permeable membrane.
This "active membrane" engulfes rods which are locally trapped in low-motility
regions and thereby further enhances the trapping efficiency by
self-organization, an effect which we call "self-encapsulation". Our results
are gained by computer simulations of self-propelled rod models confined on a 
two-dimensional planar or spherical surface with a stepwise constant motility field, 
but the phenomenon should be observable in any geometry with sufficiently large 
spatial inhomogeneity. We also discuss possibilities to verify our predictions 
of active-membrane formation
in experiments of self-propelled colloidal rods 
and vibrated granular matter.
\end{abstract}

\maketitle

\section{Introduction}
Active materials are composed of autonomously moving agents that steadily
consume energy while they are in motion. With only a small set of physical
ingredients, they can mimic the complex behavior seen in living systems, such
as swarming and flocking, directional motion, energy-fueled transport,
clustering, and bacterial turbulence
\cite{Romanczuk2012,Elgeti2015,Bechinger2016,Zottl2016}.  Over the past
decade, numerous artificial active-matter systems have been designed and
intensely studied, ranging from synthetic colloidal microswimmers on the micron
scale to self-propelled vibrated granulates on the macroscopic scale.  In many
cases, the interactions between neighbouring active particles are aligning such
that they propel towards the same direction, giving rise to a flocking effect
\cite{Vicsek1995}. A relatively new avenue of research focuses on
inhomogeneous motility fields, in which the particle self-propagation speed
depends on the spatial coordinate. This is frequently encountered in actual
biological or artificial systems where the swimming speed depends on an
external stimulus, such as an externally imposed chemical
\cite{Hong2007,Pohl,Saha2014,Liebchen1,Liebchen2,Maass}, light \cite{Drescher2010, Lozano2016} or flow
field \cite{rheotaxis1,Mathijssen2016} of the solvent.  Both linear gradients in
motility \cite{Hong2007,Nori,Lozano2016} and stepwise constant motility fields
\cite{Magiera2015} have been studied, but also more complicated motility ratchets
\cite{Stenhammar2016,Lozano2016} and even motility waves propagating in time
\cite{Geiseler1,Geiseler2,Geiseler3,Sharma}.  In general, in regions of low
motility, active particles are dynamically trapped as they move much slower
there, causing them to become locally pinned. The trapping efficiency has been
recently studied in detail for non-aligning self-propelled spheres
\cite{Magiera2015}.

\begin{figure}[t!]
  \begin{center}
    \includegraphics[width=0.42\textwidth]{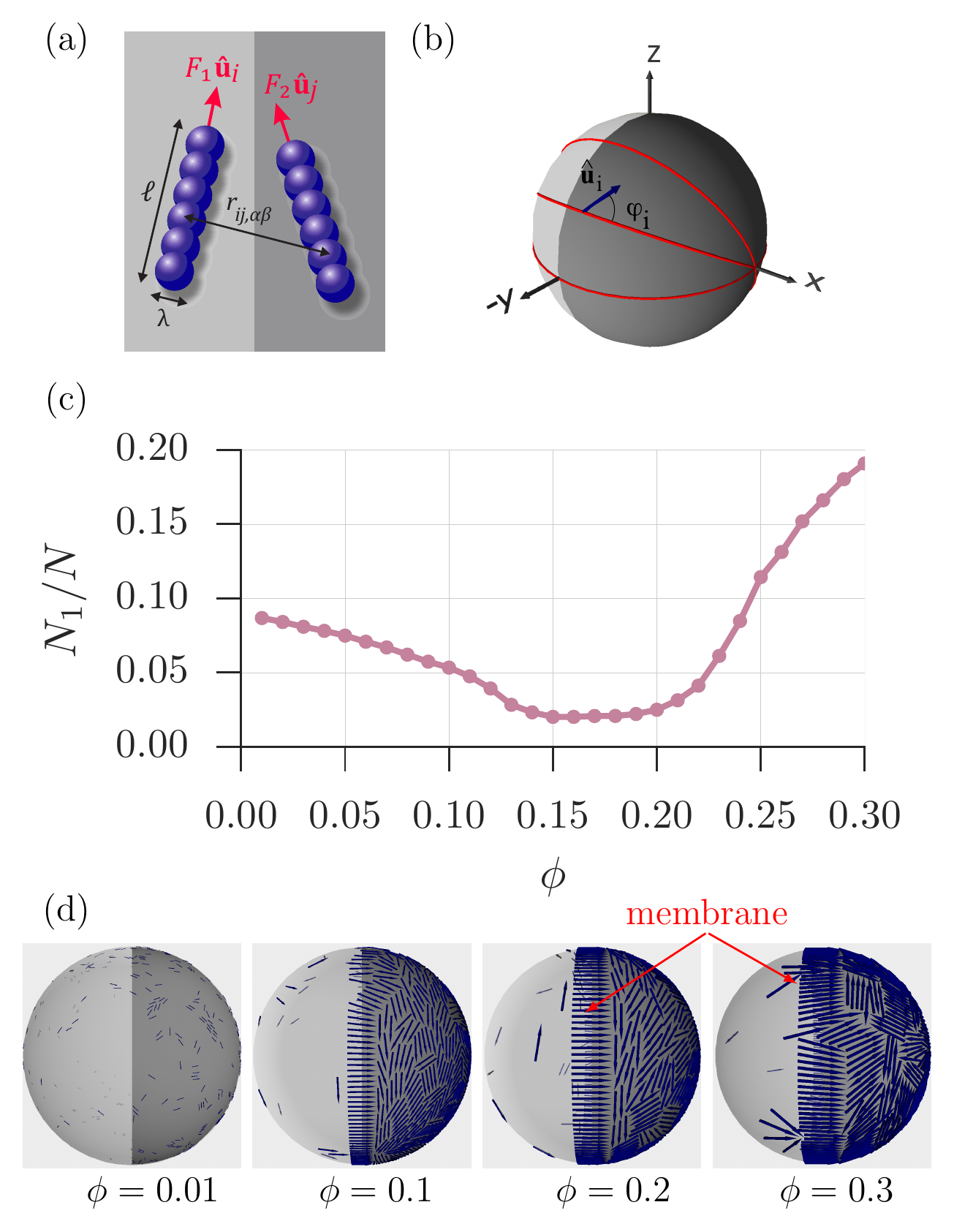}
  \end{center}
  \caption{
  \label{fig:1}
  (a) Sketch of the active-rod model. A light-gray background corresponds to a
      region with higher activity $F_1$, and dark-gray shading corresponds to
      a lower activity $F_2$.
  (b) Illustration of the coordinate system used for simulations on a spherical surface. The sphere is centered around the
      origin, and the region associated with the higher self-propulsion speed $F_1$ lies on the
      negative $x$-axis. The angle $\varphi_i$ 
      represents the angle between a rod's orientation vector $\hat{\mathbf{u}}_i$ and
      the rotated meridian perpendicular to the boundary.
  (c) Average fraction of particles $N_1/N$ residing on the "fast" $x\leq0$ hemisphere
      as a function of packing fraction $\phi$ for a system of $N=200$ rods with
      particle aspect ratio $a=10$ and activity ratio $F_1/F_2=10$.
  (d) Representative snapshots of steady-state configurations at various packing fractions,
      obtained after a total simulation time of $60000\tau$. 
  }
\end{figure}

Here we explore the behaviour of self-propelled rods with aligning interactions
in inhomogeneous motility fields that are stepwise constant. We
demonstrate that the combination of alignment and spatial inhomogeneity allows
active matter to self-organize into non-equilibrium {\it membrane-like
stuctures}, reminiscent of colloidal \cite{Barry2010,Yang2012} and biological lipid membranes
\cite{review_membranes,Andela}.
In fact, at the interface where two regions of constant but 
different motility meet, a smectic-like rod layer is spontaneously created
that acts as an effectively semi-permeable "active membrane". 
In analogy to lipid membranes, this active
self-organized structure leads to spontaneous compartmentalization, 
can be penetrated by other particles, and the number of particles
forming the membrane is not fixed but fluctuating. The active membrane
engulfes rods which are stuck in low-motility regions, an effect which we call
\textit{self-encapsulation}. Self-encapsulation can be understood as a self-organized
"fence" around trapped rods which drastically enhances the trapping efficiency,
and thus naturally leads to compartmentalization of active particles \cite{Spellings2015}.
It is worth to mention that this self-organized trapping is qualitatively
different from motility-induced phase separation, which occurs in homogeneous
motility fields \cite{Cates2015}, and from capturing self-propelled rods in wedge-like obstacles where the
trapping is induced by geometry \cite{Kaiser1,Kaiser2,Kaiser3}. Moreover the
active membranes found here are different from active nematic films driven by
anchoring and patterning \cite{Peng2016,Guillamat2016,Liebchen3}.

Our results are gained by computer simulations of self-propelled rod models
\cite{Yukawa_segment1,Wensink2008,Wensink2012,Wensink2012a} confined on a two-dimensional (2D)
surface. We consider two geometries: a spherical surface in which the region of lower motility covers one
hemisphere or a smaller surface area, and a 2D planar surface with periodic boundary conditions,
in which one half of the surface is associated with a lower motility.
We will focus mainly on the topology of the compact sphere, since it naturally gives rise to only
a single interface, and has also recently attracted interest due to its rich curvature- and topology-induced active-particle dynamics
\cite{Keber2014,Sknepnek2015,Grossmann2015,Li2015,Fily2016,Janssen2017,sphere5,sphere6}.
However, for the membrane-formation process reported here, the spherical topology
is not a crucial ingredient: indeed, we will show that self-encapsulation at the interface between two different
motilities also occurs similarly for rod motion in the plane, and that in fact the aligning interactions are
the crucial factor. Moreover, we will discuss the similarity between the 
self-encapsulation process and the growth of smectic phases out of an
isotropic phase that is impeded by rods lying perpendicular to the smectic
layer \cite{vanRoij}, an effect known as "self-poisoning" in passive systems
\cite{Schilling}. Our predictions 
may be verified in experiments of, e.g., self-propelled colloidal Janus rods
steered by external light intensity \cite{Buttinoni,Palacci2013}, active microtubuli
in a motility assay with varying kinesin motor concentrations \cite{Nitzsche2010}, and vibrated
granular matter \cite{Dauchot1}. Furthermore, our findings are of relevance for
rod-like bacteria in different motility environments \cite{Rosko2017}.

The paper is organized as follows. We first give an overview of the model systems
used in the simulations, followed by a discussion of the structural and dynamical
properties of the self-organized active membrane. We pay special attention to the
encapsulation dynamics by probing time-correlation functions of the
particle density on one side of the membrane. Lastly, we establish the robustness
of the spontaneous membrane-formation process by varying several parameters of
the motility field. We close with concluding remarks and a perspective on
possible experimental realizations of our system.

\section{Simulation model}
Our simulation model describes self-propelled particles undergoing Brownian motion
on a 2D surface with space-dependent motility. Explicitly, we have extended the 
models used by Janssen \textit{et al.}\ \cite{Janssen2017} and Wensink \textit{et al.}\ \cite{Wensink2012},
which provide a minimal description for microswimmers confined to a 
spherical and 2D planar square surface with periodic boundary conditions, respectively, to the inhomogeneous case.
In all simulations, the system is composed of $N$ rods of length $\ell$ that all experience a
space-dependent self-propulsion force along their longitudinal rod axis
$\hat{\mathbf{u}}_i$, where $i$ is the particle index [see Fig.\ \ref{fig:1}(a)]. We choose the magnitude
of the active force, $F(x_i)$, to be stepwise dependent on the Cartesian $x$-coordinate
of the rods' center-of-mass positions $\mathbf{r}_i \equiv (x_i,y_i,z_i)$ [see Fig.\ \ref{fig:1}(b)]:
\begin{eqnarray}
F(x_i) =
\begin{cases}
  F_1 & \text{if } x_i \leq 0\\
  F_2 & \text{if } x_i >0,
\end{cases}
\end{eqnarray}
with $F_1$ and $F_2$ denoting constants. For both the spherical and planar confining surface, we place the 
origin of our coordinate system in the center, so that the low- and high-motility regions comprise equal surface areas. 
The special case $F_1=F_2$ reduces to the homogeneous scenarios of Refs.\ \cite{Janssen2017,Wensink2012}. 
To account for steric repulsion among
the particles, we represent each rod as a rigid chain of
$m$ spherical segments, and let all segment-segment pairs
belonging to different rods interact through a repulsive Yukawa potential.  The
total interaction energy between two rods is given by
\begin{equation}
U_{ij} = \frac{U_0}{m^2}
\sum_{\alpha=1}^m \sum_{\beta=1}^m
\frac{\exp(-r_{ij,\alpha\beta}/\lambda)}{r_{ij,\alpha\beta}},
\end{equation}
where $U_0$ is the potential amplitude, $r_{ij,\alpha\beta}$ is the Euclidean distance
between segment $\alpha$ of rod $i$ and segment $\beta$ of rod $j$, and $\lambda$
is the screening distance that also serves as the unit of length [see Fig.\ \ref{fig:1}(a)]. 
For the spherical topology, we follow Ref.\ \cite{Janssen2017} and
constrain the rods such that $\mathbf{r}_i$ always lies on the
sphere and $\hat{\mathbf{u}}_i$ lies in the plane tangent to the sphere at
position $\mathbf{r}_i$. Within such a local tangent plane, the dynamics can
be treated as effectively two-dimensional, and hence we simulate all dynamics
by integrating the 2D overdamped Langevin (Brownian) equations of motion,
\begin{eqnarray}
  \dot{\mathbf{r}}_i &=& \bm{\mu}_T [-\nabla_{\mathbf{r}_i} U + F(x_i) \hat{\mathbf{u}}_i], \nonumber \\
\label{eq:BD}
  \dot{\hat{\mathbf{u}}}_i &=& -\bm{\mu}_R \nabla_{\hat{\mathbf{u}}_i} U,
\end{eqnarray}
where the dots denote time derivatives, $U=\frac{1}{2}\sum_{i,j\neq i} U_{ij}$,
and $\nabla_{\hat{\mathbf{u}}_i}$ is the gradient on the unit circle.  The
matrices $\bm{\mu}_T$ and $\bm{\mu}_R$ represent 
inverse translational and rotational friction tensors, 
respectively, which are defined as
\begin{eqnarray}
  \bm{\mu}_T &=& \mu_0 [\mu_{\parallel}\hat{\mathbf{u}}_i \otimes \hat{\mathbf{u}}_i
                  + \mu_{\perp}(\mathbf{I} - \hat{\mathbf{u}}_i \otimes \hat{\mathbf{u}}_i)], \\
  \bm{\mu}_R &=& \mu_0 \mu_R \mathbf{I}.
\end{eqnarray}
Here $\mu_0$ is a mobility coefficient, $\mathbf{I}$ is the $2\times2$
unit matrix, $\otimes$ stands for the dyadic product, and for the 
parameters
$\mu_{\parallel}$, $\mu_{\perp}$, and $\mu_R$ we use the standard expressions for
rod-like macromolecules as given in Ref.\ \cite{Tirado1984}.  We adopt
characteristic units such that $\lambda=1$, $\mu_0=1$, and for the unit of
activity we set $F_0=1$, so that time is measured in units of $\tau =
\lambda/(\mu_0F_0)$. For the potential we take $U_0=250$ and a cutoff distance of
$r_c=6\lambda$, and the number of segments per rod is chosen as 
$m=\lceil 14a/8 \rceil$, where $a=\ell/\lambda$ is the rod aspect ratio.
Equation (\ref{eq:BD}) is propagated using an Euler
integration scheme with a discrete time step of $0.01\tau$.  For
simplicity we have ignored any stochastic noise and hydrodynamic interactions 
(HI)--implying that the dynamics is governed solely by the repulsive pair interactions and
self-propulsion forces--, but we have verified that the membrane is also
stable against small thermal noise, as will be shown below. Due to the neglect of HI, our 
model is particularly
suitable for dry active matter, but we note that HI-free simulations can also 
accurately reproduce the complex behavior seen in hydrodynamic models, including
active turbulence \cite{Wensink2012} and compartmentalization of active spinners \cite{Spellings2015}.

\section{Membrane formation and structure}
We first explore the emergent structural and dynamical properties as a function of the
packing fraction $\phi$, defined for the spherical surface as $\phi = N\ell \lambda/(4\pi R^2)$, where $R$ is the radius of
the confining sphere, and for the planar surface as  $\phi = N\ell \lambda / \ell_{\rm{box}}^2$, where 
$\ell_{\rm{box}}$ is the width of the square simulation box.
Let us first focus on the spherical-surface case. 
Figure \ref{fig:1}(c) shows the fraction of rods on the
high-motility $x\leq0$ hemisphere, $N_1/N$, as a function of $\phi$ for a system of $N=200$
rods with aspect ratio $a=10$ and self-propulsion strengths
$F_1=1F_0$ and $F_2=0.1F_0$. In the dilute limit of $\phi\rightarrow 0$,
all rods behave as free particles that spend a fraction $F_2/(F_1+F_2)$
of the time on the left hemisphere, implying $N_1/N=0.09$, which indeed we
observe numerically. As the density increases up to $\phi\approx 0.2$, however,
we find a remarkable effect: the "fast" region with $x\leq0$ becomes depleted and an
excessive amount of particles will reside at the hemisphere with the lower
self-propulsion speed $F_2$. The reason for this becomes evident from the
particle snapshots, Fig.\ \ref{fig:1}(d): at the boundary between the two
hemispheres, particles self-organize into a \textit{membrane-like structure}
that effectively prevents particles from leaving the "slow" $x>0$ region, thus acting
as a self-encapsulation mechanism. The formation of this membrane arises from three
crucial ingredients: i) the spatial inhomogeneity of the motility, $F_1>F_2$, which naturally
imposes an inhomogeneous density profile, ii) a sufficiently high packing
fraction, which allows for saturation of rods on the $x>0$ hemisphere, and iii)
aligning interactions, which emerge from pair collisions between the active rods.
Indeed, we have verified that the membrane
structure disappears if $F_1 \approx F_2$, $\phi \ll 0.2$, or $a \ll 10$.
We also note that the formation of the membrane is fostered by the
periodicity of the sphere: if a rod is able to permeate
through the interface and move into the $x\leq0$ region, it will swim
across the entire hemisphere and eventually collide with membrane-forming
particles on the other side, consequently causing it to align and becoming part of the
membrane itself. Through a similar mechanism, we see that for higher packing
fractions $\phi>0.2$, where the "slow" $x>0$ hemisphere is fully saturated with
particles, ''hedgehog" structures \cite{Wensink2008} appear on the "fast" $x\leq 0$ side of the
membrane. Thus, a polar ordering of particles oriented toward the domain associated with
lower motility emerges naturally near the interface.
\begin{figure}[t!]
        \begin{center}
    \includegraphics[width=0.32\textwidth]{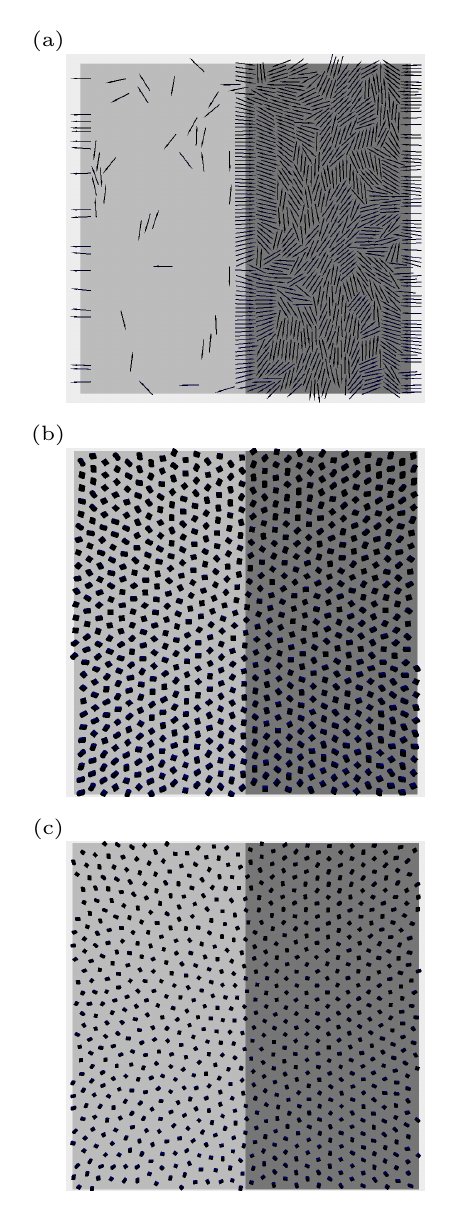}
  \end{center}
  \caption{
  \label{fig:2}
  Representative snapshots of steady-state configurations for $N=800$ particles and activity ratio $F_1/F_2 = 10$
  on a 2D surface with periodic boundary conditions.
  (a) Particle aspect ratio $a=16$ and 
  packing fraction $\phi=0.2$, (b) particle aspect ratio $a=1$ and packing fraction $\phi=0.2$, with  $\mu_{\parallel}=\mu_R=1$ and $Pe=100$,
  and (c) particle aspect ratio $a=1$ and packing fraction $\phi=0.1$, with $\mu_{\parallel}=\mu_R=1$ and $Pe=100$.
  In the latter two cases, the particles are spherically shaped and experience no aligning torques during collision.
  }
\end{figure}

For rods residing on a flat 2D surface with periodic boundary conditions, the observed behavior is very similar to that on the sphere:
at packing fractions $\phi \approx 0.2$ a clear membrane structure appears at the interface
between the regions of different motilities. Due to the periodic boundary conditions, we
now find two separate membranes at $x=0$ and $|x|=\ell_{\rm{box}}/2$ which encapsulate the rods in the low-motility region from opposing sides.
Figure \ref{fig:2}(a) shows a typical snapshot of this scenario for $N=800$ rods with aspect ratio $a=16$ and $F_1/F_2 = 10$.
As in the spherical case, a membrane is formed only when the rods are sufficiently elongated to induce sufficiently strong aligning pair collisions,
and indeed the membrane structure becomes increasingly distorted as the rod aspect ratio decreases. 

To unambiguously confirm that aligning interactions are crucial, we have also performed 2D calculations for spherical particles with $a=1$ 
that experience no torque during collision. In this case, particle reorientation may only occur through rotational Brownian diffusion. In order
to account for such diffusional motion and thus to allow for a fair comparison between the dynamics of rods and spheres, we have extended our simulation model 
to finite P\'{e}clet number $Pe = \mu_0 F_1/\sqrt{D_{\parallel} D_R}$, 
where $D_{\parallel}=\mu_0 \mu_{\parallel} k_{\rm{B}}T$ and $D_R = \mu_0 \mu_R k_{\rm{B}}T$ are translational and rotational diffusion coefficients,
respectively, $k_{\rm{B}}$ is the Boltzmann constant, and $T$ is a temperature. 
For elongated rods, we have verified that the membrane structure is robust against thermal Brownian translational and rotational noise.
For spherical particles, however, we find a markedly different pattern: in the
absence of explicitly aligning interactions, the particles form an active
crystalline phase at $\phi=0.2$ that covers the entire surface homogeneously.
Figure \ref{fig:2}(b) shows a snapshot of such a phase for $N=800$, $a=1$,
$\mu_{\parallel}=\mu_R=1$, $Pe=100$, and $F_1/F_2=10$. Here the average number of particles
is the same in the $x<0$ and $x>0$ regions, and consequently the lattice constants are 
identical for the high- and low-motility domains.
We have checked that the
observed crystalline pattern for this packing fraction also appears for lower $Pe$ values and higher
particle numbers (up to $N=4000$), thus ruling out possible finite-size effects.
As the packing fraction decreases to $\phi=0.1$, however, the high-motility region becomes fluidized,
as shown in Fig.\ \ref{fig:2}(c). Thus, rather than forming a membrane, the motility edge for spherical particles becomes a
fluid-crystal interface, with the fluid on the high-activity and the hexagonal crystal on the low-activity side.
This reveals that, at the appropriate density range, one can steer fluid and crystal phases at wish by inhomogeneous motility fields.
We note that the formation of a hexagonally ordered phase of active Brownian
particles in a low-motility region has also been reported by Magiera and Brendel \cite{Magiera2015}
for spherical particles with a shorter-ranged repulsive interaction potential.
The detailed Brownian dynamics of spherical active particles in 
inhomogeneous motility fields will be discussed in a separate publication; for
the present work, we only emphasize that aligning torques are necessary to form
the here-reported membrane structure.

\begin{figure*}[t!]
        \begin{center}
    \includegraphics[width=0.75\textwidth]{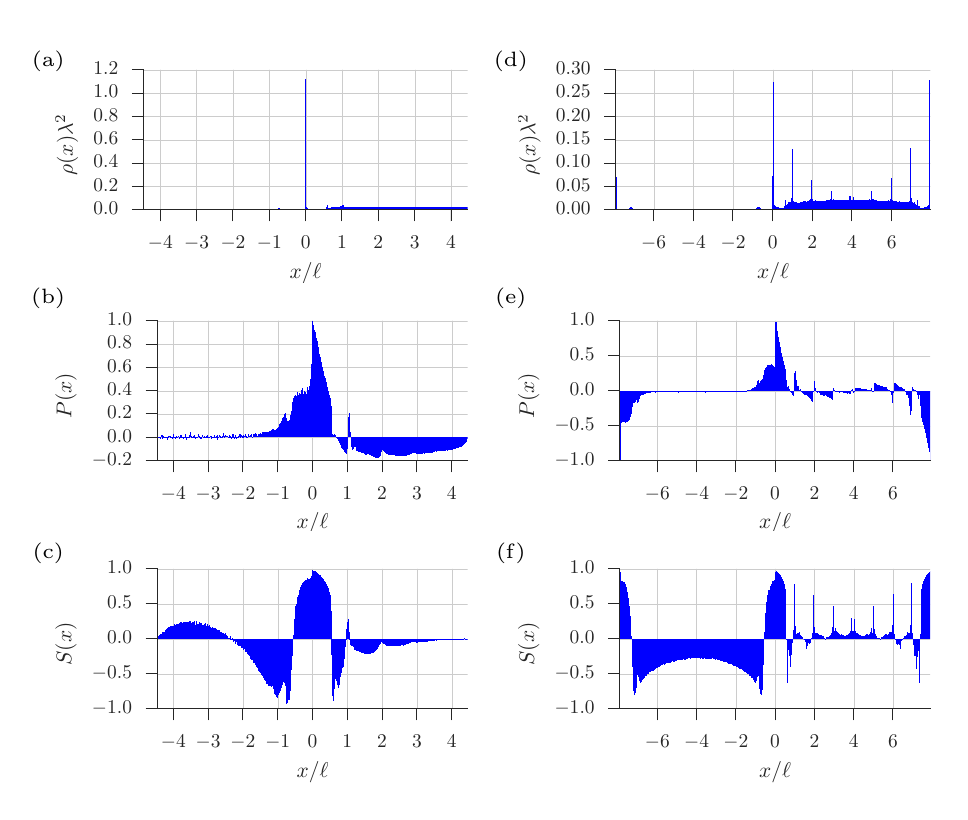}
  \end{center}
  \caption{
  \label{fig:3order}
  Structural order parameters for $N=800$ active rods with aspect ratio $a=16$ at packing fraction $\phi=0.2$ and 
  activity ratio $F_1/F_2=10$, calculated for (a-c) a spherical confining surface, and
  (d-f) a 2D surface with periodic boundary conditions.
  (a,d) Average density profile $\rho(x)\lambda^2$, (b,e) polar order parameter $P(x)$, and 
  (c,f) nematic order parameter $S(x)$. In all panels, the $x$-coordinate is normalized by the rod length $\ell=16\lambda$.
  }
\end{figure*}
In order to characterize the structure of the self-organized membrane of active rods, we
calculate the density profile $\rho(x)$, the polar order parameter $P(x)$, and the nematic
order parameter $S(x)$, which are defined as
\begin{eqnarray}
\rho(x) &=&  \sum_{i=1}^N \langle \delta(x-x_i) \rangle, \\   
P(x) &=& \frac{1}{N} \sum_{i=1}^N \left\langle \cos(\varphi_i) \right\rangle_x, \\ 
S(x) &=& \frac{1}{N} \sum_{i=1}^N \left\langle 2\cos^{2}(\varphi_i) -1 \right\rangle_x,
\end{eqnarray}
where $\langle \cdot \rangle$ denotes an ensemble average and $\langle \cdot \rangle_x$ an average under the constraint that the position of 
the $i$-th rod is at $x_i=x$.
For the spherical confining surface, $\varphi_i$ is the angle between the rod axis $\hat{\mathbf{u}}_i$ and rotated meridian perpendicular to
the boundary [see Fig.\ \ref{fig:1}(b)], while for the 2D plane $\varphi_i$ represents the angle between $\hat{\mathbf{u}}_i$ and the
positive $x$-axis.
In both cases, the values of $P(x)$ and $S(x)$ can range from $+1$ to $-1$.
Figure \ref{fig:3order} depicts the $\rho(x)$, $P(x)$, and $S(x)$ profiles for systems with a fully developed membrane 
on a spherical and planar 2D surface, respectively, all calculated for $N=800$, $\phi=0.2$, and $F_1/F_2 = 10$. 
The data are averaged over 30 and 50 independent trajectories, respectively. 

The average density profiles, Fig.\ \ref{fig:3order}(a) and (d), indicate that almost all particles reside on the region with
low motility, $x>0$, confirming a high trapping efficiency on both the spherical and
planar surface. Moreover, in both cases, the membrane at $x\approx0$ is
composed of a large number of particles 
with
nearly perfect parallel alignment along the $x$-axis, since $P(0)\approx1$ [Fig.\ \ref{fig:3order}(b,e)] 
and $S(0)\approx1$ [Fig.\ \ref{fig:3order}(c,f)]. 
Note that by symmetry, the membrane at $|x|=\ell_{\rm{box}}/2$ on the planar surface has 
the opposite polarization.
Furthermore, in the 2D planar case, a second, third, and even fourth row of parallel-oriented particles is clearly 
visible at $x/\ell \simeq 1, 2$ and 3, respectively. For the spherical surface, however, this smectic ordering is 
frustrated by the curvature of the sphere, and only one additional layer of particles is apparent at $x/\ell\simeq 1$.
We have verified that the smectic layering on the sphere is further enhanced when the
inhomogeneity ratio increases, $F_1 \gg F_2$. 

Curiously, on both sides of the
membrane, at $x/\ell\approx-0.8$ and $x/\ell\approx0.8$, 
we find a subset of particles aligned \textit{perpendicular} to the membrane-forming rods,
as evidenced by the locally strongly negative nematic order parameter $S(x)$. 
 Note
that this occurs similarly for the spherical and planar case.
The origin of the transverse order in the high-density (i.e.\ low-motility)
region is, however, qualitatively different from that occuring in the
low-density region. In the high-density region,
intralayer particles with perpendicular ordering arise from packing:
this effect
occurs already in bulk equilibrium \cite{vanRoij,Bartsch2017} and was also found
as a "self-poisoning" scenario for crystallizing hard-rod
liquids by Schilling and Frenkel \cite{Schilling}. The
perpendicularly oriented particles effectively hamper permeation
of rods across the membrane, thus partially shielding the
self-organized structure. On the low-density side, perpendicular
ordering has mainly a dynamical origin. Particles which are perpendicular just
repeatedly move around more times
contributing therefore more strongly to the average (both in the circular
and planar situation). As the distance from the interface increases,
the structural ordering becomes less distinct. For the
spherical surface, this means that at the poles with $|x| = R$,
the order parameters $S(x)$ and $P(x)$ both level off to zero due to symmetry.
It may be seen in Fig.\ \ref{fig:3order}(c) that, when approaching the $x=-R$ pole of the low-density region, 
the nematic order parameter $S$ changes again sign at $x/\ell \simeq -2.3$.
This is due to the definition
of the latitudinal reference orientation, as a
typical trajectory nearby the pole will be on average
 more likely to be parallel than perpendicular
to the latitudinal direction.  For
the 2D planar case, the rods moving in the high-motility region
($x < 0$) have a weak propensity to remain perpendicular
to the membrane-forming rods, leading to a net negative value
of the nematic order parameter even in the middle between the
two opposing membranes, at $x =-\ell_{box}/4$; this is
attributed to the finite size of the simulation box, which we have verified by comparing simulations for
$N=800$ and $N=2000$. The net polarization
is negligible however, $P(-\ell_{box}/4)\approx 0$,
which naturally follows from the absence of any symmetry-breaking mechanism along the $y$-axis.
Overall, we can conclude that the structure and encapsulation function of the membrane, which forms at
the interface between a high- and low-motility region, is qualitatively similar for a spherical and planar surface,
but the details of the particle ordering away from the membrane depend on the geometry and symmetry of the imposed confinement.

\section{Permeation dynamics}
We next turn our attention to the permeation dynamics of the membrane and monitor how
the concentration of particles in the high-motility region ($x\leq 0$) 
evolves as a function of time. For simplicity we focus here on the spherical-surface case, since 
it gives rise to only a single membrane, but all results can be assumed to be qualitatively similar to the 
2D planar case.
We define an incoherent time-correlation function
probing the \textit{single-particle} dynamics,
\begin{equation}
\label{eq:Cs}
C_s(t) = \frac{1}{N} \sum_{i=1}^N \frac{\langle n_i(0) n_i(t)\rangle}{\langle n_i^2 \rangle},
\end{equation}
where $n_i$ measures on which hemisphere a particle resides,
\begin{eqnarray}
n_i =
\begin{cases}
  1 & \text{if } x_i \leq 0\\
  0 & \text{if } x_i >0,
\end{cases}
\end{eqnarray}
and a coherent time-correlation function probing the \textit{collective} dynamics, 
\begin{equation}
\label{eq:Cc}
C_c(t) = \frac{\langle N_1(0) N_1(t)\rangle}{\langle N_1^2 \rangle},
\end{equation}
where $N_1 = \sum_{i=1}^N n_i$. 
Furthermore, we also consider the average particle flux per unit time and unit length, $J/(\tau \lambda)$, which 
measures the net number of particles crossing the interface towards the high-motility region.
In all cases, the average is taken over a maximum of $700$ independent trajectories
with a total simulation time up to $1.2 \cdot 10^6\tau$ each.  In order to
expedite the statistical averaging process, we here focus on smaller systems
with $N=200$, but we have verified that the qualitative picture applies also
for larger system sizes of $\mathcal{O}(1000)$.
\begin{figure}[h!]
        \begin{center}
    \includegraphics[width=0.39\textwidth]{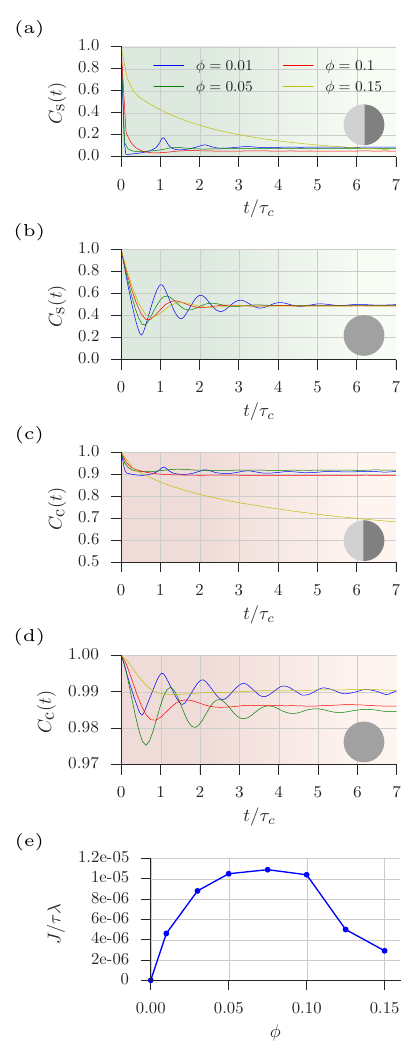}
  \end{center}
  \caption{
  \label{fig:3}
  (a-d) Time-correlation functions calculated for $N=200$ rods with $a=10$
  at different packing fractions $\phi$. The top two panels show the incoherent $C_s(t)$
  functions [Eq.\ (\ref{eq:Cs})] for (a) inhomogeneous activity
  ratio $F_1/F_2=10$ and (b) \textit{homogeneous} activity, i.e.\ $F_1/F_2=1$.
  Insets illustrate the motility field.
  The lower panels show the corresponding collective $C_c(t)$ 
  functions [Eq.\ (\ref{eq:Cc})] for (c) $F_1/F_2=10$ and (d) $F_1/F_2=1$.  Time
  is given in units of the characteristic time $\tau_c$ in which a free particle
  will swim across one great circle of the sphere. All legends are as in panel (a).
  Panel (e) shows the average particle flux $J/(\tau \lambda)$ across the interface for $F_1/F_2=10$ as a function of packing fraction $\phi$.
  }
\end{figure}

As a reference case, let us first consider the dynamics in the dilute limit 
where particles behave as free swimmers and no membrane structure is formed.
In this scenario, every rod will swim independently across a great circle of the sphere,
spending time periods of relative duration $F_2/(F_1+F_2)$ and $F_1/(F_1+F_2)$ on the
$x\leq 0$ and $x>0$ hemispheres, respectively.
The corresponding canonically averaged time-correlation function $C_s(t)$ will,
after a brief initial decay, oscillate around the average value $\langle n_i^2 \rangle \equiv \langle n_i \rangle = F_2/(F_1+F_2)
\approx 0.09$
with a period of $\tau_c$, i.e., the time it takes a free particle to cover one
great circle. Our simulations at $\phi=0.01$ numerically confirm this picture,
as can be seen from Fig.\ \ref{fig:3}(a). While in this case all particles may cross
the interface at $x=0$ without experiencing any steric hindrance, the normalized particle flux $J/(\tau \lambda)$
is still close to zero, as shown in Fig. \ref{fig:3}(e). This is simply due to the very 
low particle density and correspondingly large interface length, resulting in an almost negligible
flux per unit of length.

At a slightly higher packing fraction of
$\phi=0.05$, we witness the onset of membrane formation: particle collisions
promote the formation of polar domains and rods accumulate on the region with
lower activity.  While not forming a fully developed membrane across the entire
sphere, but rather a dynamic polar structure that is constantly dissolved and
rebuilt locally, the membrane-like domains do transiently trap particles and
effectively delay the crossing time between the two hemispheres.  Indeed, the
long-time value $\langle n_i \rangle$ is slightly lower than in the
free-particle case, and importantly, the oscillation period of $C_s(t)$ at
$\phi=0.05$ is a factor of 1.5 \textit{larger} than the time $\tau_c$ expected
for non-interacting particles.  To unambiguously establish that this is not
merely a result of the increased particle density, we also compare our results
against the \textit{homogeneous} case with uniform activity
$F_1=F_2={2\over11}\,F_0$, in which case the characteristic time $\tau_c$ is
identical but the membrane is absent.  Figure \ref{fig:3}(b) reveals that this
scenario would result in a $C_s(t)$ oscillation period of approximately 1.2
times $\tau_c$, thus confirming that the presence of the membrane delays the
dynamics and gives rise to enhanced trapping and self-encapsulation.  This
trend continues as the packing fraction further increases to $\phi=0.1$, in
which case the homogeneous reference scenario reveals oscillations in $C_s(t)$
with a period of $1.5 \tau_c$, while the inhomogeneous $C_s(t)$ data show only
a single oscillation and decay to a lower long-time value, indicative of the
fact that particles become more strongly trapped behind the membrane on the
$x>0$ hemisphere. Figure \ref{fig:3}(e) also shows that the net flux 
slightly decreases at $\phi=0.1$, consistent with an enhanced trapping effect
due to the partially permeable membrane.

Finally, at a packing fraction of $\phi=0.15$, a "perfect" membrane is formed
that covers the entire interface at $x=0$, and almost all remaining particles
are encapsulated in the densely-packed $x>0$ region. Here the membrane has
completely lost its permeability and instead acts as a self-organized trapping
barrier that fosters a maximal accumulation of rods on one side of the sphere.
A comparison of the time-correlation functions in panels \ref{fig:3}(a) and
\ref{fig:3}(b) for $\phi=0.15$ confirms this picture: in the inhomogeneous
($F_1/F_2=10$) case, there is not a single oscillation in $C_s(t)$ visible on
the time scale considered in this work, and instead we observe only a very slow
decay pattern in which a limited number of particles manages to change
hemispheres. Conversely, the homogeneous ($F_1/F_2=1$) case exhibits an
oscillatory pattern with a period of $1.8 \tau_c$, implying that here particles
can depart and re-enter the "slow" $x>0$ hemisphere far more easily. From Fig.\
\ref{fig:3}(c) we also see that the coherent time-correlation function $C_c(t)$
at $\phi=0.15$ decays to a much lower long-time value than at lower packing
fractions, indicating that the particle number $N_1$ exhibits far greater
fluctuations. This is due to the fact that $N_1$ is minimal when the
encapsulation effect is maximal [cf.\ Fig.\ \ref{fig:1}(a)], implying that
\textit{any}  particle crossing the interface at $\phi=0.15$ will constitute a
relatively large change in $N_1$, and thus a relatively strong decorrelation in
$C_c(t)$.  We also note that for all packing fractions considered here, the
permeation dynamics is governed predominantly by \textit{single-particle}
crossing events, rather than groups of collectively  permeating particles, and
indeed the decay of $C_c(t)$ is enslaved by $C_s(t)$.

As a final confirmation of the strong trapping effect at $\phi=0.15$, 
we observe a distinct drop in the particle flux at this packing fraction 
[Fig.\ \ref{fig:3}(e)]. Collectively, these results thus unambiguously show 
that the self-organized membrane structure leads to high trapping efficiency 
and slow permeation dynamics, resulting in the spontaneous compartmentalization of particles.
We have verified for P\'{e}clet numbers of $Pe=100$ and 300 that the inclusion of thermal translational
and rotational Brownian noise does not alter this qualitative picture, 
but does lead to a higher effective permeability.


\section{Dependence on geometric parameters}
\begin{figure}
        \begin{center}
    \includegraphics[width=0.36\textwidth]{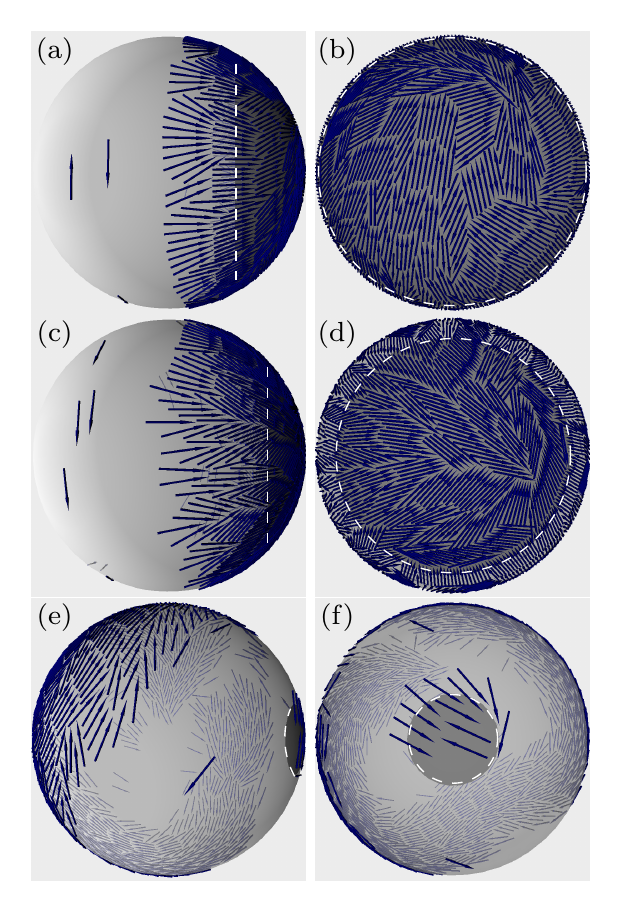}
  \end{center}
  \caption{
  \label{fig:4}
  Representative snapshots for a system of $N=800$ rods with aspect ratio $a=16$ 
  and packing fraction $\phi=0.2$, with step locations of the motility field located at 
  (a,b) $x_0 = 0.55R$, (c,d) $x_0=0.78R$, and (e,f) $x_0=0.97R$, as indicated by
  the white dashed lines. Left panels show the ($x,z$) plane and right panels the ($y,z$) plane.
  }
\end{figure}

Let us finally investigate how robust the spontaneous membrane-formation process is
against variations in the geometric parameters of the motility field, namely the surface area
of the region associated with lower activity, and the inhomogeneity ratio $F_1/F_2$.
To vary the surface area we consider the case
\begin{eqnarray}
F(x_i) =
\begin{cases}
  F_1 & \text{if } x_i \leq x_0\\
  F_2 & \text{if } x_i > x_0
\end{cases}
\end{eqnarray}
with $x_0>0$, so that the region of lower self-propulsion speed will become
increasingly small as $x_0$ increases. Figure \ref{fig:4} shows representative
shapshots for a system of $N=800$ rods on a spherical surface with $\phi=0.2$ and $F_1/F_2=10$, for
different interface locations $x_0 = 0.55R, 0.78R$, and $0.97R$. We find  
that up to $x_0 \approx 0.9R$, a membrane-like smectic ordering is consistently formed across the interface, but
becomes more distorted as the low-motility region becomes smaller. For the packing
fraction considered here, the surface area available in the low-motility domain 
is insufficient to accommodate all particles, even for small values of $x_0>0$.
Hence, 
particles must inevitably reside in the region with higher activity, forming
distorted smectic layers and hedgehog-like structures around the $F_2$
domain--similar to what we found for the $x_0=0$ case at higher packing
fractions. We thus conclude that the low-motility region acts as a "nucleation"
core around which particles accumulate, even as this domain becomes completely
saturated with particles. This phenomenon is akin to the behavior reported earlier
for non-aligning active spheres in inhomogeneous media \cite{Magiera2015}, and implies that even
a small inhomogeneity can effectively trap an excessive number of particles.
Consequently, this behavior may readily be exploited in applications that require
spatial control over active particles. In the limit of $x_0 \rightarrow R$, i.e.,
a vanishing low-motility region, we recover the homogeneous scenario of Ref.\
\cite{Janssen2017} which, for the rod aspect ratios and densities considered
here, gives rise to a flocking state [cf.\ Fig.\ \ref{fig:4}(e,f)].

\begin{figure}
        \begin{center}
    \includegraphics[width=0.4\textwidth]{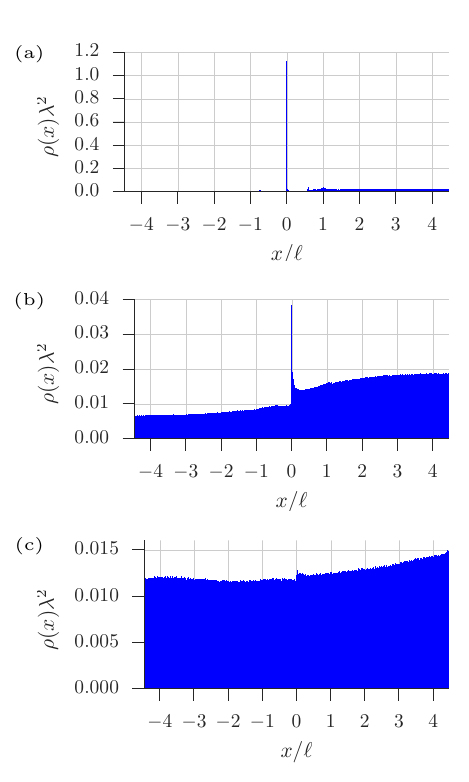}
  \end{center}
  \caption{
  \label{fig:5}
  Density profiles $\rho(x)\lambda^2$ for a system of $N=800$ rods with aspect ratio $a=16$,
  packing fraction $\phi=0.2$, and interface location $x_0=0$, 
  for different activity ratios $F_1/F_2$ of (a) 10, (b) 2, and (c) 1.1.
  }
\end{figure}
Lastly, by varying the ratio between self-propulsion speeds, $F_1/F_2$, we
confirm that the membrane structure is fostered by a strong inhomogeneity.
Figure \ref{fig:5} shows the dimensionless density profiles $\rho(x)\lambda^2$ for a spherical surface with $N=800$,
$\phi=0.2$, and $x_0=0$ with activity ratios $F_1/F_2 = 10, 2$, and
$1.1$. It is clear that a larger difference in motilities leads to a
more prominent membrane structure at the interface. As the values of $F_1$ and
$F_2$ approach each other, the rods experience a smaller difference in
self-propulsion speed and the density profile becomes more homogeneous across
the entire sphere. In the limit of $F_1=F_2$ we again recover the flocking
state of Ref.\ \cite{Janssen2017} for the spherical surface, and of Ref.\ \cite{Wensink2012} for the planar 2D surface. 
This unambiguously confirms that spatial
inhomogeneity is a crucial ingredient for the membrane formation and
self-encapsulation of active rods.

%

\section{Conclusions}
In conclusion, we have established a link between the physics of membranes and
self-propelled particles: in an inhomogeneous motility field, an {\it active
membrane\/} is spontaneously formed by a competition between self-propulsion
and rod interactions. The effect is robust and occurs in any geometry
provided there is a steep jump in motility over the rod length. The active
membrane encapsulates particles trapped in a low-motility region and
significantly enhances the trapping efficiency. This possesses applications to
capture and steer microswimmers efficiently via motility fields.

In principle it is possible to verify our predictions in experiments. One
feasible realization consists of colloidal Janus rods driven by light
\cite{Buttinoni,Palacci2013}, which can be exposed to almost arbitrary motility
landscapes \cite{Lozano2016}.  Similar possibilities exist for self-propelled
droplets \cite{Maass} or modular microswimmers steered by an electrolyte
gradient \cite{Palberg}. Moreover, rod-like bacteria at high concentrations
\cite{Wensink2012,Rosko2017} may serve as another model system to observe
smectic ordering in motility landscapes.  Lastly, macroscopic rod-like granulates
can be made active by vertical vibration \cite{Junot2017,Dauchot1} and different
motilities can in principle be controlled by imposing an upper frictional zone
parallel to the vibrating table. This kind of "dry" active matter is in particular
an excellent realization of our model as hydrodynamic interactions are absent. 
For the future it would be interesting to also incorporate solvent-mediated
hydrodynamic interactions between the rods into our model. Due to the
constrained motion of the rods on the spherical surface this will be a highly
nontrivial task.

Finally, there is an increasing interest in microscopic statistical theories for
interacting active particles. Dynamical density functional theory
\cite{Wittmann,ourDDFT1,Wensink2008,ourDDFT3} is an appropriate approach to obtain
predictions for layering of rod-like particles.  In order to describe the
effects theoretically, these theories need to be supplemented and generalized
towards an inhomogeneous motility field, see Ref.\ \cite{Sharma} for a recent
study in this direction.

\acknowledgments
H.L. acknowledges the DFG for support through project LO 418/20-1.
L.C.M.J. thanks the Alexander von Humboldt Foundation for support through a
Humboldt Research Fellowship.

%

\end{document}